\documentclass[12pt]{article}
\usepackage{amsmath,amsfonts,amssymb,latexsym}
\usepackage{epsfig}
\usepackage{graphicx}
\usepackage{wrapfig}
\usepackage{subfigure}
\usepackage{cite}

\newcommand{\sect}[1]{\setcounter{equation}{0}\section{#1}}

\def\be{\begin{equation}}
\def\ee{\end{equation}}
\def\bea{\begin{eqnarray}}
\def\eea{\end{eqnarray}}
\def\bean{\begin{eqnarray*}}
\def\eean{\end{eqnarray*}}

\def\Z{{\mathbb Z}}
\def\R{{\mathbb R}}

\begin{document}


\thispagestyle{empty}
\hfill \today

\vspace{2.5cm}

\begin{center}
\bf{\LARGE 
Algebraic special functions and so(3,2) 
}
\end{center}

\bigskip\bigskip

\begin{center}
E. Celeghini$^1$  and M.A. del Olmo$^2$
\end{center}

\begin{center}
$^1${\sl Departimento di Fisica, Universit\`a  di Firenze and
INFN--Sezione di
Firenze \\
I50019 Sesto Fiorentino,  Firenze, Italy}\\
\medskip

$^2${\sl Departamento de F\'{\i}sica Te\'orica, Universidad de
Valladolid, \\ 
E-47011, Valladolid, Spain.}\\
\medskip

{e-mail: celeghini@fi.infn.it, olmo@fta.uva.es}

\end{center}

\bigskip

\bigskip

\begin{abstract}

A ladder structure of operators is presented for the associated Legendre polynomials and the spherical harmonics showing that both belong to the same  irreducible representation of $so(3,2)$. As both are  also bases of square-integrable functions, the  universal enveloping algebra of $so(3,2)$ is thus  shown to be isomorphic to the space of linear operators acting on the $L^2$   functions defined on   $(-1,1)\times \Z$ and on the sphere $S^2$, respectively.   

The presence of a ladder structure is suggested to be the general condition to obtain a Lie algebra representation defining in this way the  ``algebraic special functions'' that are proposed to be the connection between Lie algebras and square-integrable functions so that the space of linear operators on the $L^2$ functions is isomorphic to the universal enveloping algebra.

\end{abstract}

\vskip 1cm


\vfill\eject


\sect{Introduction}\label{intro}

Almost all functions with a name have been sometimes included in the category of  special functions. Indeed 
 ``special'' functions as opposed to the ``standard''  ones could be defined  as the functions  that are relevant for applications such that  someone has given a name to them \cite{berry2001}. In this approach ``special'' is simply a synonymous of useful.

Following Talman~\cite{talman1968} and Truesdell~\cite{truesdell1948} we are restricting here to  special functions that belong  to a wide but not so inclusive class. Talman relates special functions to representations of elementary groups, showing that in some cases many  of their properties have a group representation origin, while Truesdell introduces the idea of ``familiar special functions'' as a subclass related to a  set of  
formal properties.
 
 We consider thus a set of formal properties in a similar way of Truesdell but taking into account Talman's ideas too.
 Special functions we consider  depend from one or more continuous variables and one or more discrete variables,  they satisfy ordinary linear differential equations of second order and this equation  can be factorized by means of ladder operators. In particular,  Hermite, Laguerre and Legendre polynomials verify the above properties  allowing    to prove that: (i) they are unitary irreducible representations (UIR) of  a rank one simple Lie algebra, (ii) they determine an orthogonal complete basis of square-integrable functions  defined on the same domain and (iii), because of (i) and (ii), the space of linear operators acting on these ${L}^2$  functions is isomorphic to the universal enveloping algebra (UEA) of the corresponding Lie algebra 
  \cite{celeghini2012}. 
  
  The goal of this paper is to show that these properties are not restricted to the above mentioned one-dimensional cases since the  associated Legendre polynomials (ALP) and the spherical harmonics (SH) satisfy the same conditions and verify the same relations (i)-(iii)  with an underlying   Lie algebra of rank two. Thus, a  subclass of special functions,  that we call ``algebraic special functions'', can be introduced  satisfying the above mentioned conditions and properties. 

The paper is organized as follows. 
Section~\ref{alp} is devoted to present the main properties of  the ALP relevant for our discussion.  
In section~\ref{so21section}   we consider the ALP with fixed $m$ and with $l\geq |m|$, 
that  allows us to 
build up by using of a set of ladder operators  
a family of UIR of  $so(2,1)$ labelled by the parameter $m$. 
In section~\ref{so3section}  we construct, by means of a different set of ladder operators  with fixed $l$  the $so(3)$ UIR of dimension $2 l+1$, well known  in the case of the spherical harmonics 
$Y_{l}^{m}(x)$. 
In section~\ref{so32section}, starting from the results of the previous sections, the whole  Lie algebra 
$so(3,2)$ and the UIR supported by the complete set of  ALP  
are constructed. 
Section~\ref{sphericalharmonicssection} is devoted to
extend the preceeding results to the SH:
like for the ALP the defining equation of the SH 
is related to the Laplace-Beltrami equation of $so(3,2)$ obtained from the differential realization of its second order Casimir and of all its three-dimensional subalgebras.
In section~\ref{scalprod} the isomorphism between the space of the operators on the $L^2$ functions 
and the  UEA of $so(3,2)$ is discussed. 
Finally, in section~\ref{conclusionssection}, some conclusions and comments  are included.

\sect{Associated Legendre polynomials: a selective view}\label{alp}

The starting point of the discussion is
the factorization method \cite{schrodinger,infeld-hull1951}  that relates the   second order differential equations defining the special functions to the   recurrence formulae with first 
order derivatives.
However, the fundamental limitation of this approach is  
that the problem has been considered  from the point of view of differential equations where the indices 
(normally only one) are considered as parameters \cite{miller1968}. The dependence of the formulas from the 
indices in iterated applications must thus be introduced by hand.

As proposed in  \cite{celeghini2012} 
a consistent vector space framework
-where the indices are related to operators- 
allows to define recurrence formulas in an operator way and
is more effective in the description of the underlying mathematical structure.

Hermite, Laguerre and Legendre polynomials have been thus  
 consistently reformulated, showing that they are a basis of a well 
defined unitary representation of a  Lie algebra of rank one.
At the same time, these functions are a basis
of the $L^2$  functions defined on the appropriate open
connected subsets of the real line \cite{Cambianis}. These $L^2$ functions belong thus to the same Lie algebra
representation such that the space of linear operators acting on them is isomorphic to the UEA
(i.e., to the space that has as a basis the ordered monomials constructed on the generators) \cite{celeghini2012}. 
Recurrence formulae are related with the algebra generators, while the defining  
second order differential equation is obtained as a second order Casimir operator $C_2$ of the symmetry  algebra.

If these properties can be generalized to all those objects having a ladder structure,  a global class of familiar
special functions could be considered, where the defining differential equation depends on as 
many parameters as the rank of the underlying algebra.

In this paper we show that at least  two relevant 2-dimensional generalizations are possible:    the  SH and 
the  ALP  (for brevity we consider in detail only the ALP and extend  at the end the discussion to the SH). Both are found to be  UIR of the real form $so(3,2)$ of the rank two  Lie algebra $B_2$ with $C_2 = -5/4$. 

The ALP we consider are  solutions of the general Legendre 
equation
\be\label{legendreeq}
(1-x^2) \; y''-2 x\, y'+\left( l(l+1)-\frac{m^2}{1-x^2}\right)\; y=0 ,
\ee
 with integers (degree) $l\in \Z^{\geq 0}$ and (order) $m\in \Z,$\;  with $ 0\leq |m|\leq l$ 
and $x\in (-1,1)$. 
The extension from  the interval $(-1,1)$  to a generic interval $(a,b) \subset \R$ is trivial and will not be discussed here.

Because the ALP are defined with different conventions by different authors, we adopt here the ones 
of \cite{NIST}, i.e.
with $m\ge 0$:
\be\label{le}
P_l^m(x)\; \; =\; (-1)^m (1-x^2)^{m/2} \frac{d^m}{dx^m} P_l(x)\;,\quad\quad 
P_l^{-m} = (-1)^m\; \frac{(l-m)!}{(l+m)!} P_l^m(x)\,,
\ee
where the $P_l(x)$ are the Legendre polynomials.
Many recurrence relations are supported by the ALP suggesting the presence of a hidden 
large symmetry\cite{abramowitz,NIST,thooft}. 
Our starting point is the following set of formulae that include first order derivatives:
\be\begin{array}{lll}
(x^2-1)\; P_l^{' m}(x)&=& \sqrt{1-x^2} \; P_{l}^{m+1}(x)+ m\, x\; 
P_{l}^{m }(x), \label{recurrence1}
\\[0.3cm]
(x^2-1)\; P_l^{' m}(x)&=&-( l+m) ( l-m+1) \sqrt{1-x^2}\; P_l^{m-1}(x)- m\, x\; P_{l}^{m }(x),
\\[0.3cm]
 (x^2-1)\; P_l^{' m}(x)&=& -( l+1)\,  x\, P_l^{m}(x)+(l-m+1) \; P_{l+1}^{m}(x),
\\[0.3cm]
 (x^2-1)\; P_l^{' m}(x)&=&  l\, x\; P_l^{m}(x)-(l+m)\;  P_{l-1}^{m}(x).
\end{array}\ee

While eq.~(\ref{legendreeq}) depends from $m^2$\, only, 
eqs.~(\ref{le}) exhibit a different normalization of the $\{P_l^m(x)\}$ for\, $m>0$\, and\, $m<0$\,. 
As all the twentieth century physics has shown, symmetry is fundamental in understanding phenomena. Hence, we eliminate this asymmetric normalization 
rescaling the $P_{l}^{m}(x)$,
in partial agreement with the usual normalization of spherical harmonics,  defining:
\be\label{Tpolynomials}
T_l^m(x):= \sqrt{\frac{(l-m)!}{(l+m)!} }\; P_{l}^{m}(x) 
\ee
that will be the basic objects of this paper. 
This change is essential to highlight the Lie algebra structure hidden behind (\ref{recurrence1}).  The elements $T_l^{m}(x)$  \,and\, 
$T_l^{-m}(x)$\, are equal up to a phase in the configuration space, as
\[
T_l^{-m}(x) = (-1)^m\; T_l^{m}(x)\;,
\] 
but --as discussed in the following-- they are eigenvectors of different eigenvalues of the 
Hermitian operator $M$ 
and thus different vectors in the vector space  $\{ T_l^{m}(x)\}$.
Note that, for $m=0$, $T_l^m(x)$ --as well as $P_l^m(x)$-- coincide with the Legendre polynomials $P_l(x)$\,.

In terms of $\{T_l^m(x)\}$ orthogonality and completeness of ALP for fixed $m$ are \cite{thooft}
\be\label{orthtm}
 \int_{-1}^{1}  T_{l}^{m}(x)\;  \left( l+ 1/2\right )
\; T_{l'}^{m}(x) \;dx= \delta_{l\,l'}\qquad (l,l' \geq |m|),
\ee
\be\label{comptm}
\sum_{l=|m|}^\infty  T_l^m(x) \left(l+1/2\right) T_l^m(y) = \delta(x-y).
\ee

Let us now display the complete operator structure on the set 
$\{T_l^m(x)\}$  we need.
In consistency with the quantum theory approach, we have to introduce not only the 
operators $X$ and $D_X$ of the configuration space, such that 
\[
X\, f(x) = x\, f(x) ,\qquad D_x\, f(x) = f'(x),\qquad [X,D_x]= -1\, ,
\]
but  also two other operators $L$ and $M$ such that 
\be\label{nnn}
L\; T_l^m(x) = l\; T_l^m(x)\,,\qquad M\; T_l^m(x) = \, m\; T_l^m(x) . 
\ee
These operators $L$ and $M$ allow to take into account the changes in the value of the parameters $m$ and $l$ after  subsequent applications of eqs.~(\ref{recurrence1})  
 (see Ref.~\cite{celeghini2012}).
Thus, the operators 
\bea\label{nJmasmenosoperators}
J_\pm &:=& \mp\sqrt{1-X^2} \;D_x-\frac{ X}{\sqrt{1-X^2} }\;M,
\\[0.3cm] \label{nKmasoperators}
K_+ &:=& -(1-X^2) \;D_x+  X\;( L+1),
\\[0.3cm]\label{nKmenosoperators}
K_- &:=& (1-X^2\,)\;D_x +  X L 
\eea
give a formal expression of the 
eqs.~(\ref{recurrence1})  
rewritten as:
\bea
 J_+\; T_l^m(x)\, &=&\, \sqrt{(l-m)(l+m+1)}\; T_l^{m+1}(x)  \label{rec1},
\\[0.3cm]
J_-\; T_l^m(x) &=&\, \sqrt{(l+m)(l-m+1)}\; T_l^{m-1}(x)  \label{rec2},
\\[0.3cm]
K_+\; T_l^m(x) &=&\, \sqrt{(l-m+1)(l+m+1)}\; T_{l+1}^{m}(x)  \label{rec3},
\\[0.3cm]
K_-\; T_l^m(x) &=&\, \sqrt{(l+m)(l-m)}\; T_{l-1}^{m}(x)  \label{rec4}.
\eea
As shown in next sections,
consecutive applications of the operators $J_\pm$ and $K_\pm$
(that in the course of the procedure modify the values of the parameters $l$ and $m$) 
allow to recover the equation \eqref{legendreeq}.
The  introduction of the operator $L$ and $M$ in 
(\ref{nJmasmenosoperators})--(\ref{nKmenosoperators}) is  
 irrelevant in eqs.~(\ref{rec1})--(\ref{rec4}) since they are  diagonal on the $T_l^m(x)$ and  we have taken care to write them always to the right 
  but, as they
do not commute with $J_\pm$ and $K_\pm$,   a more complex algebraic scheme appears:
\[
[M,L]=0,\quad [M, J_\pm ] = \pm J_\pm, \quad [M, K_\pm ] =0, \quad [L,J_\pm]=0, \quad [L,K_\pm]= \pm K_\pm .
\]

The main result of this paper  is that factorization is only the first step of
a more formal algebraic approach where 
eq.~\eqref{legendreeq} can be reconstructed  not only
as   product of  operators but also as  
 Casimir operator  of a    Lie algebra.
Indeed, starting from the operators $J_\pm$ and $K_\pm$ and their commutators, 
we obtain a Lie algebra representation of $so(3,2)$ with 
$C_2 = -5/4$. The operators
 $L$ and $M$, above introduced,  belong to
the Cartan subalgebra  and the
ALP (as well as the SH) are a basis of this UIR. 

The relevance of this result is related 
from one side to the role of intertwining between differential equations and Lie algebras played
by these algebraic special functions
and from the other to the fact that the UEA (i.e., the vector space that has as a basis the ordered monomials
of the generators) of $so(3,2)$ is isomorphic to the space of the linear operators
acting on the corresponding space of  square-integrable functions. 


\sect{so(2,1)  and associated 
Legendre polynomials}\label{so21section}

In this section we prove that the set of APL\; $T_{l}^{m}(x)$  with fixed\, $m$\, and   
 $ l\in \Z^{ \geq |m|}$ 
supports a $m$-dependent   UIR  of the discrete series of $so(2,1)$.

First of all let us consider, starting from eqs.~(\ref{rec3})--(\ref{rec4}),    the factorization approach.
The two operators  $K_\pm$ act on the $\{T_l^m(x)\}$ by rising and lowering the label $l$ keaping fix the label $m$. 
Taking into account that the operator $L$ \eqref{nnn} takes different values on 
$T_l^m(x)$ and  $T_{l\pm 1}^m(x)$, 
recurrence relations allow to recover eq.~(\ref{legendreeq}) by 
reiterated application. Indeed remembering 
eqs.~\eqref{nKmasoperators} and \eqref{nKmenosoperators},  the relations
\[
K_+ K_-\;\, T_l^m(x) = (l^2-m^2) \; T_l^m(x)\;, \qquad K_- K_+\;\, T_l^m(x) = ((l+1)^2-m^2) \; T_l^m(x)\;,
\]  
allow to write
\be\label{rr1}
\, -(1-X^2)\;\left( (1-X^2) D_x^2  -2 X D_x + L(L+1)-\frac{1}{1-X^2}\;M^2\right)\equiv 0 ,
\ee
that, disregarding the irrelevant factor $-(1-X^2)$,  never zero in the domain $(-1,+1)$, is the 
operator form of eq.(\ref{legendreeq}).

Alternatively we can  introduce the Lie algebra generated by $K_\pm$. 
The commutator and the anticommutator of $K_\pm$ act on the space $\{T_l^m(x)\}$ as
\bea \label{KmasKmenscom}
[K_+,K_-]\; T_l^m(x)&=&- 2\left(l+1/2\right)\; \; T_l^m(x), \\[0.3cm]
\{K_+,K_-\}\; T_l^m(x)&=&2  \left(l(l+1)-m^2+1/2\right) \;T_l^m(x).
\nonumber\label{KmenosJmasanticom}
\eea
Defining, as suggested by eq.~\eqref{KmasKmenscom},
\be\label{k3}
K_3:= L+ 1/2
\ee 
we  see that $K_3$ togheter
with $K_\pm$ close a $so(2,1)$ Lie algebra 
\be\label{Kso21comm}
[K_3,K_\pm]= \pm  K_\pm, \qquad
[K_+,K_-]=- 2 K_3 ,
\ee
with Casimir operator related to $m$: 
\be\label{so21casimir}
{\cal C}_2 \; T_l^m(x)\equiv \left(K_3^2-\frac 12 \{K_+,K_-\}\right) \;\, 
T_l^m(x)= (m^2-1/4)\; T_l^m(x)\;.
\ee
 As $m=0, \pm1, \pm2 \dots$ 
the double valued  discrete  UIR of $so(2,1)$ are obtained
 with
${\cal C}_2\equiv m^2-1/4=\,- 1/4,\, 3/4,\, 15/4,\dots$ \cite{Bargmann}. 
As $l\geq |m|$  the spectrum of the operator 
$K_3$ is related to the parameter $m$, and has the eigenvalues 
$ 1/2 +|m|, 3/2 +|m|, 5/2 +|m|,\dots$  

\begin{figure}[h]
\centerline{\psfig{figure=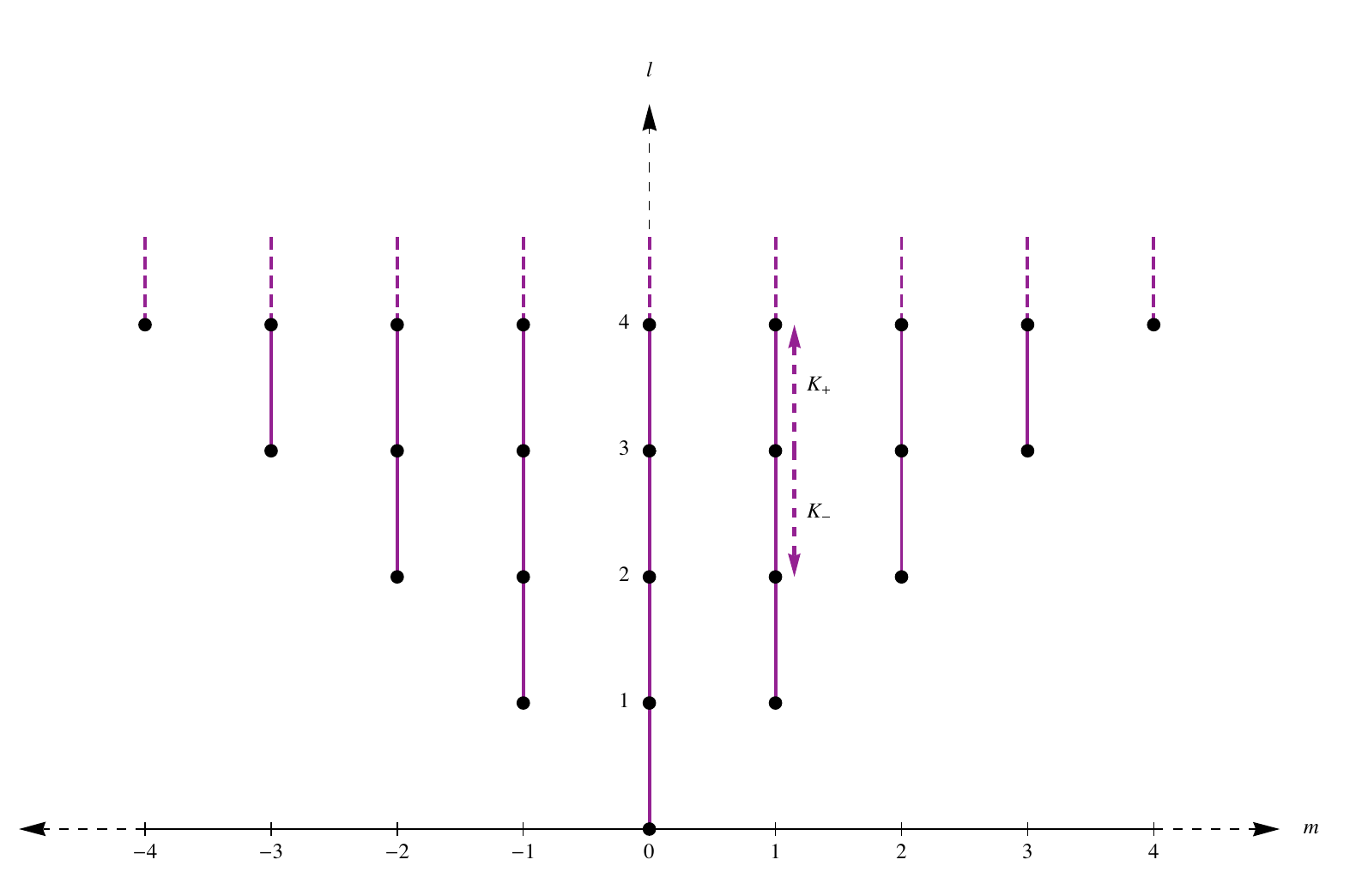, height=8.0cm}}
\caption{\small Classification of the ALP\,  $T_{l}^{m}(x)$  (black points) in terms of  the 
UIR,  $U^m$, of $so(2,1)$
labeled by  $m$ (vertical lines).  The    operator 
$K_3$ is diagonal with eigenvalue $l+1/2$. The action of the operators $K^\pm$ on $T_3^1(x)$ is  also displayed.  Note that
 $U^m$ and $U^{-m}$ are equivalent.} \label{fig_2}
\end{figure}
Summarizing,  the set of ALP\, $T_l^m(x)$  with any fixed $m\in \Z$ and 
$ l=|m|, |m|+1, |m|+2, \dots$ 
supports the double valued UIR, $U^m$, of $so(2,1)$ of the discrete series  (Figure~\ref{fig_2}). 

Taking now into account the differential representation of the operators  \eqref{nKmasoperators} and 
\eqref{nKmenosoperators}  we see also that the Casimir 
\eqref{so21casimir}  of the $so(2,1)$ algebra reproduces eq.~(\ref{rr1}), 
i.e., up to the irrelevant global factor $(1-X^2)$, the operator form of the 
generalized Legendre equation \eqref{legendreeq}:
\[\label{so21casimireq}
{\cal C}_2- (M^2-1/4)=
\, (1-X^2)\;\left( (1-X^2) D_x^2-2 X D_x + L(L+1)-\frac{1}{1-X^2}\;M^2\right)\equiv 0 .
\]

\sect{so(3)   and associated Legendre polynomials}\label{so3section}

Let us now consider the structure related with the operators $J_\pm$. They do not change the value of\, $l$
but they are, as shown by eqs.~(\ref{rec1})--(\ref{rec2}),  the well known rising and lowering operators of the 
rotation algebra $so(3)$ in the representation $l$.

Again the Legendre equation can be easily recovered from the recurrence relations. 
Indeed, we have from eqs.~(\ref{rec1}) and  (\ref{rec2})
\be\label{JmasJmenos}
J_\pm J_\mp\;T_l^m(x) =
(l\pm m)(l\mp m + 1)\; T_l^m(x),
\ee
that in the  differential representation 
\eqref{nJmasmenosoperators}  is written
\be\label{dJmasJmenos}
J_\pm J_\mp\;T_l^m(x) =
\left( -(1-X^2) D_x^2 + 2 X D_x + \frac{X^2}{1-X^2}\; M^2\pm M\right)\; T_l^m(x)\;. 
\ee
Comparing  expressions  \eqref{JmasJmenos} and  \eqref{dJmasJmenos}, the operator form of the 
generalized Legendre equation \eqref{legendreeq} is easy recovered 
\[\label{dJmasJmenoseq}
 (1-X^2) D_x^2 - 2 X D_x + L(L+1)-\frac{1}{1-X^2} M^2 \; = 0 .
\]
\begin{figure}[h]
\centerline{\psfig{figure=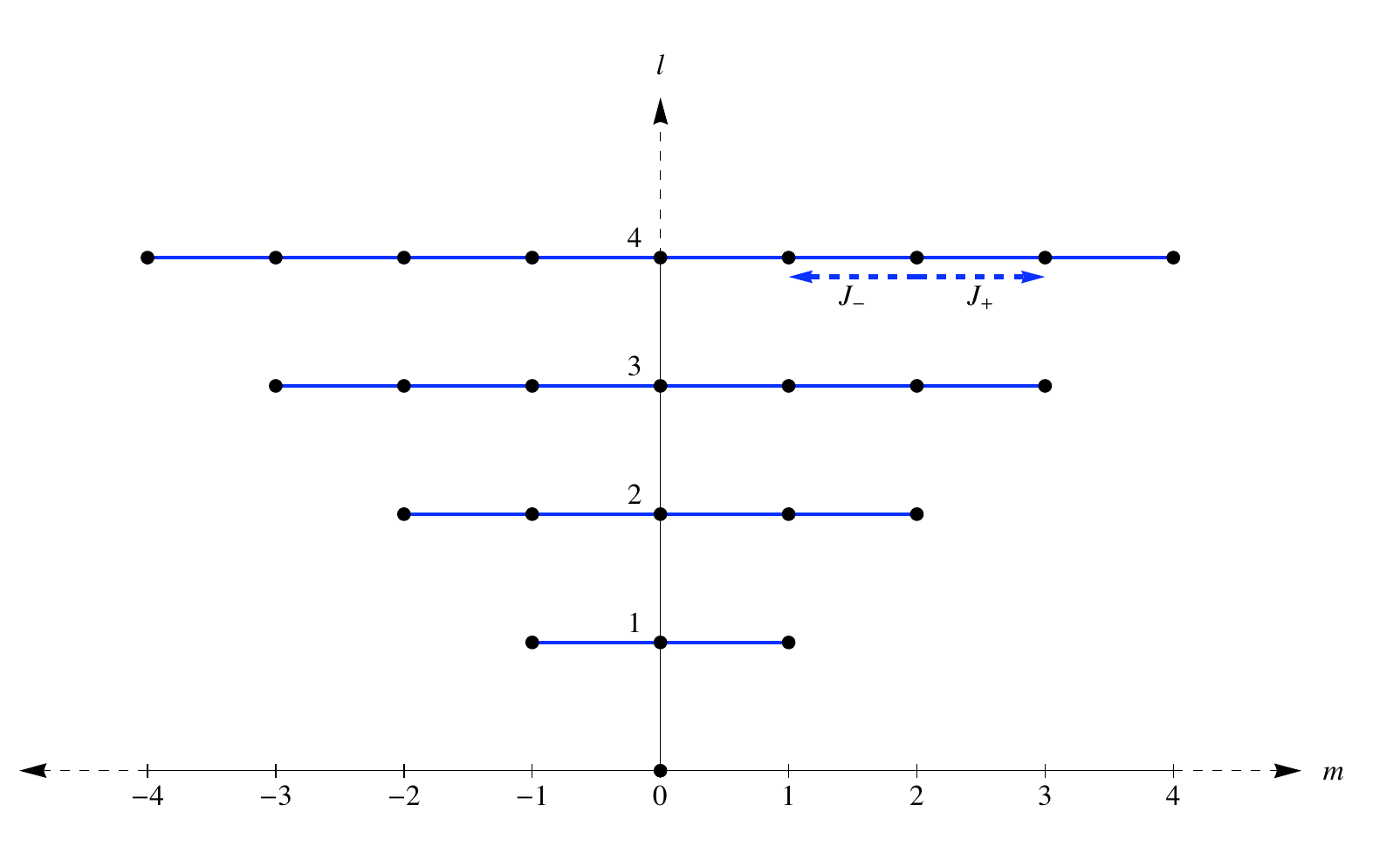, height=8.0cm}}
\caption{\small Classification of the ALP  $T_{l}^{m}(x)$ (black points) in terms of  the IUR,  $U^l$, of  $so(3)$ (horizontal segments).  The operator $J_3$ is diagonal with eigenvalue $m$.  The action of the operators 
$J_\pm$ on $T_4^2(x)$ is  displayed.} \label{fig_1}
\end{figure}

As in the preceeding section, let us now introduce the Lie algebraic approach.
From eqs.~(\ref{JmasJmenos}) commutator and anticommutator are
\bea \label{JmasJmenscom}
[J_+,J_-]\;  T_l^m(x)&=& 2\, m\; T_l^m(x),\nonumber \\[0.3cm]
\{J_+,J_-\}\;  T_l^m(x)&=& 2 \left( l(l+1)-m^2\right) \; T_l^m(x) ,
\nonumber\label{JmenosJmasanticom}
\eea
and, defining 
\be\label{Jtres}
J_3:= M ,
\ee
we get  the $so(3)$ algebra
\be\label{so3comm}
[J_3,J_\pm]= \pm  J_\pm\,, \quad\qquad
[J_+,J_-]= 2 J_3\; 
\ee
with Casimir 
\[
{\cal C}_2 \;   T_l^m(x)=\left(J_3^2+\frac 12 \{J_+,J_-\}\right) \; 
 T_l^m(x) =\, l(l+1)\; T_l^m(x)\,. 
\]
Starting now from the differential representation of the operators 
$J_\pm$ \eqref{nJmasmenosoperators},  we can  rewrite 
\[
{\cal C}_2  -L(L+1)
=-\left( (1-X^2) D_x^2 - 2 X D_x + L(L+1)-\frac{1}{1-X^2}\;M^2\right)
\equiv 0 ,
\]
 i.e., the operator equation \eqref{legendreeq} now obtained  from the Casimir of the subalgebra $so(3)$ with fixed 
 $l$.

Summarizing,  the set of ALP $T_l^m(x)$  with any fixed  $l\in \Z$ and 
$ m=-l,-l+1,\dots, l-1,l$\; supports the  unitary irreducible  representation $U^l$ of $so(3)$, 
as shown in the Figure~\ref{fig_1}.  


\sect{so(3,2) and associated 
Legendre polynomials}\label{so32section}

Starting from the operators  
(\ref{nJmasmenosoperators})--(\ref{nKmenosoperators}) and their actions  (\ref{rec1})--(\ref{rec4})
on the vector space of the APL $T_l^m(x)$,
we  generalize to the full vector space $\{T_l^m(x)\}$ the
algebraic structures discussed in sections \ref{so21section} and \ref{so3section}. 

Indeed, as both for $m$ or $l$ fixed we have found an underlying   algebra, we can look for a global Lie algebra
of which  both $so(2,1)$ and $so(3)$ are subalgebras,
such that the set of all the ALP $\{T_l^m(x)\}$  supports a representation of this  algebra.
As the operators $K_+$ and  $J_\pm$ applied 
to $T_0^0(x)=1$ generate the full space $\{T_l^m(x)\}$ 
\[
T_l^{\pm m}(x)= \frac{1}{l!}  \;\sqrt{\frac{(l-m)!}{(l+m)!}}\;(J_\pm)^{m}\;(K_+)^l \;T_0^0(x), \qquad m\geq 0\,,
\]
hence the representation, if it exists, is irreducible.

Since $\langle K_3, K_\pm\rangle$ and $\langle J_3, J_\pm\rangle$ should be subalgebras, we have to consider only the mixed commutators.
So, defining
\[
R_\pm := [K_\pm,J_\pm]
\]
 we obtain from  (\ref{rec1})--(\ref{rec4})
\bea\label{accionRmenos}
 && R_+\;T_l^m(x)=
\sqrt{(l+ m+2)(l+ m+1)}\; T_{l+1}^{m+1}(x),
\\[0.3cm] \label{accionRmas}
&& R_-\;T_l^m(x)=
\sqrt{(l+ m)(l+ m-1)}\;  T_{l-1}^{m-1}(x),
\eea
and from eqs.~(\ref{nJmasmenosoperators})--(\ref{nKmenosoperators}) the differential form of the operators  $R_\pm$
\bea\label{nRmasoperators}
&& R_+= -X \;\sqrt{1-X^2}\;D_x -\frac{1}{\sqrt{1-X^2}}\; M -\sqrt{1-X^2}\; (L+1) ,\\[0.3cm]
&& R_-= X\; \sqrt{1-X^2}\;D_x -\frac{1}{\sqrt{1-X^2}}\; M -\sqrt{1-X^2} \; L  . \label{nRmenosoperators}
\eea
In a similar way to the previous cases the  general Legendre equation
\eqref{legendreeq} can be obtained by means of the factorization method applying the recurrence relations $R_\pm R_\mp$ to $T_l^m(x)$.
\begin{figure}
\centerline{\psfig{figure=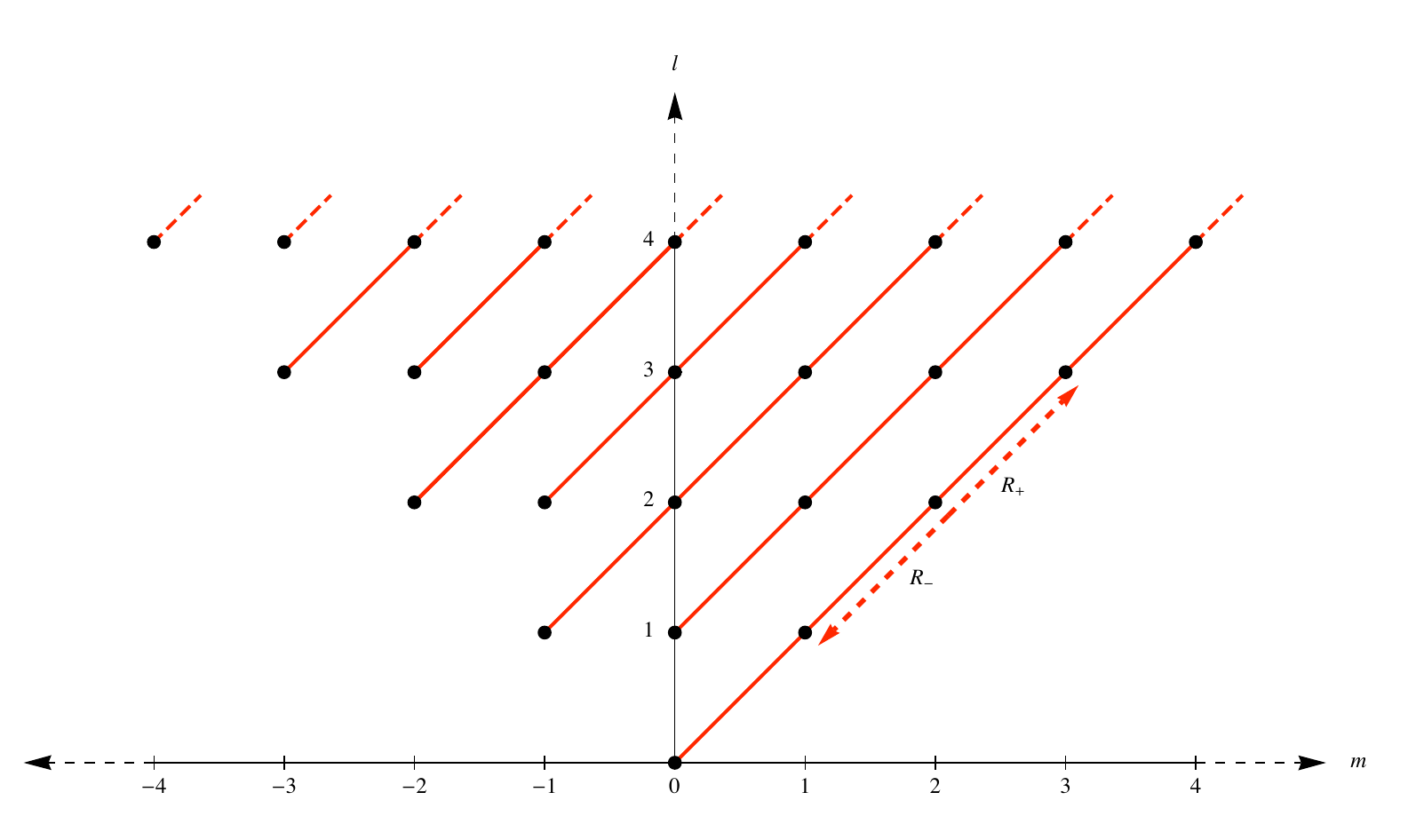,height=8.0cm}}
\caption{\small Classification of the ALP  $T_{l}^{m}(x)$  (black points) in terms of  the UIR of   $so(2,1)\equiv \langle R_\pm,R_3\rangle$ (inclined lines). The operator $R_3$ is diagonal with eigenvalues 
$l+m+1/2$. The action of the operators 
$R^\pm$ on $T_2^2(x)$ is also displayed.} \label{fig_3}
\end{figure}

Moreover, as
\be \label{Rso21comm}
[R_+,R_-]=-4 (L+M+1/2) ,\qquad  
\ee
we can define $R_3:= L+M+1/2$ and find
that $R_\pm$ and $R_3$ span a $so(2,1)$ algebra
\be\label{commR}
[R_+,R_-]=- 4 R_3, \qquad [R_3,R_\pm]=\pm 2 R_\pm \, .
\ee
The
Casimir gives 
\[
R_3^2-\frac 12 \{R_+,R_-\} +\frac 34\equiv 0
\]
that, written in terms of the differential form \eqref{nRmasoperators} and \eqref{nRmenosoperators} of $R_\pm$
reproduces again, up to an irrelevant $X^2$  
global factor,
the operator   Legendre equation \eqref{legendreeq} 

\[\label{so21casimireqR}
R_3^2-\frac 12 \{R_+,R_-\} +\frac 34\, \equiv\,  
 X^2\left( (1-X^2) D_x^2  -2 X D_x +L(L+1)-\frac{1}{1-X^2}\;M^2\right)= 0,
\]
that can be thus derived also from the subalgebra 
$\langle R_\pm, R_3 \rangle$, i.e.,  $so(2,1)$ with $l-m$ fixed.
Moreover, rescaling the generators as
\[ 
R_i \to R'_i=\frac 12 R_i \, ,
\]
eqs.~(\ref{commR}) reproduce
the standard form for the Lie commutators of $so(2,1)$ as reported in eq.~\eqref{Kso21comm} and thus
for all these (infinite) representations 
of $so(2,1)$ the Casimir ${\cal C}_2$ is $-3/16$. Since  
${\cal C}_2= k(k-1)$ the maximal weight is $k=1/4$ or $3/4$.  The spectrum of $R_3$   
is indeed\; $1/2,5/2,9/2, \dots$\, when\, $l+m$\, is even and\; $3/2,7/2,11/2, \dots$ when $l+m$  is odd 
(see Figure~\ref{fig_3}). 

In a similar way   
we define two other new operators,   denoted by 
$S_+$ and $S_- ,$
\[
S_\pm := [K_\pm,J_\mp].
\]
Their actions  on the ALP $T_{l}^{m}(x)$ are  
\bea\label{accionSmenos}
 && S_+\;T_l^m(x)=
\sqrt{(l-m+2)(l- m+1)} \; T_{l+1}^{m-1}(x),
\\[0.3cm] \label{accionSmas}
&& S_-\;T_l^m(x)=
\sqrt{(l- m)(l- m-1)} \;  T_{l-1}^{m+1}(x)
\eea
and their differential forms
\bea\label{nSmasoperators}\nonumber
&& S_+= X \sqrt{1-X^2}\;D_x -\frac{1}{\sqrt{1-X^2}}\; M +\sqrt{1-X^2}\; (L+1) ,\\[0.3cm]
&& S_-= -X \sqrt{1-X^2}\;D_x -\frac{1}{\sqrt{1-X^2}}\; M +\sqrt{1-X^2} \; L  . \nonumber
\eea
\begin{figure}
\centerline{\psfig{figure=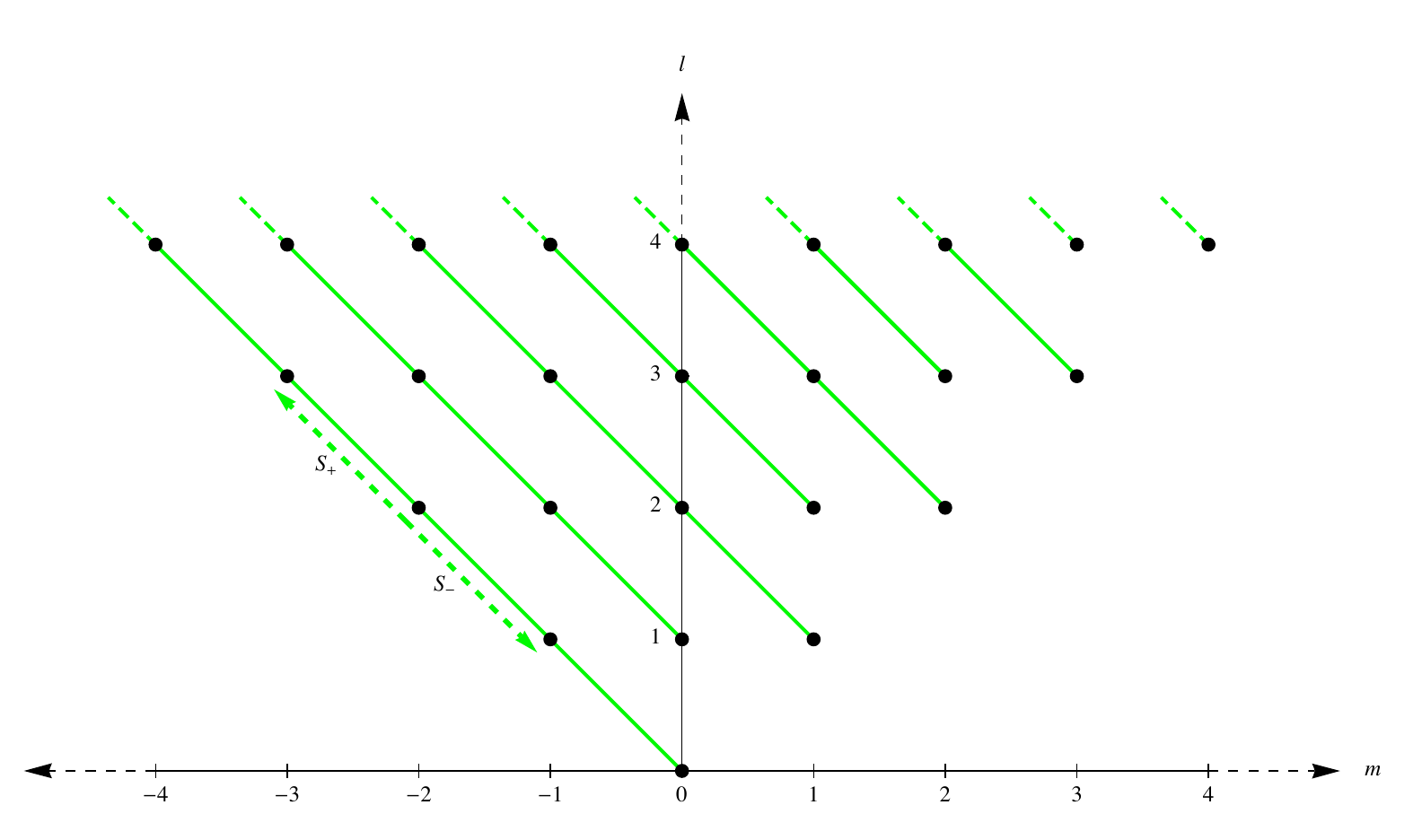,height=8.0cm}}
\caption{\small Classification of the ALP  $T_{l}^{m}(x)$  (black points) in terms of  the UIR of  
$so(2,1)\equiv \langle S_\pm,S_3\rangle$ (inclined lines). The operator $S_3$ is diagonal with eigenvalues $l-m+1/2$. The action of  
$S^\pm$ on $T_2^{-2}(x)$ is also displayed.} \label{fig_4}
\end{figure}

Again, like before, the Legendre equation can be obtained applying 
the recurrence relations $ S_\pm S_\mp$ to  $T_l^m(x)$.
The situation is similar to the previous case also for the algebraic approach: in analogy with \eqref{commR},
defining $S_3:= L-M+1/2$, we have
 \be\label{Sso21comm}
 [S_+,S_-]=-4 S_3 , \qquad
[S_3, S_\pm ] = \pm 2 S_\pm .
\ee
By inspection, $\langle S_\pm,S_3\rangle $ is a new $so(2,1)$ algebra obtained from   $\langle R_\pm,R_3\rangle$ 
with  the substitution
$M\to -M$.  It has thus  the same representation  
 with Casimir ${\cal C}_2 = -3/16$
than $\langle R_\pm,R_3\rangle$ 
(see Figures \ref{fig_3} and \ref{fig_4}) and allows  to obtain again
the   Legendre equation \eqref{legendreeq}.

\begin{figure}
\centerline{\psfig{figure=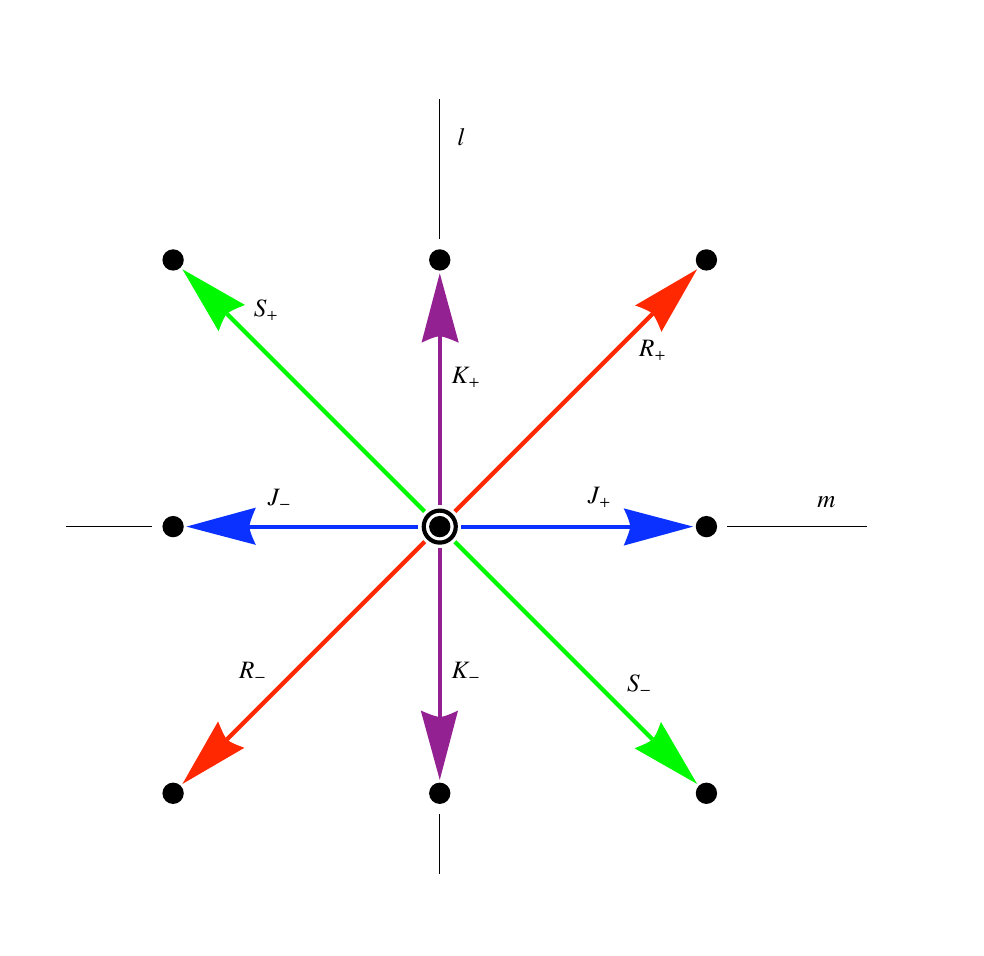,height=8.0cm}}
\caption{\small Root system of $B_2$. The two-dimensional Cartan subalgebra is represented at the origin.} \label{fig_5}
\end{figure}

So, we have four set of  operators $\{J_\pm,J_3\}$,  $\{K_\pm, K_3\}$, \{$R_\pm,R_3\}$, and $\{S_\pm,S_3\}$ 
generating the first one a $so(3)$ algebra and each of the other three a $so(2,1)$ algebra.  
All together  these operators close a bigger Lie algebra containing all these four algebras as subalgebras. 
First of all notice that 
\[
R_3=J_3+K_3,\qquad S_3=-J_3+K_3 ,
\]
hence, the   rank of this algebra  is two as only two elements
of the Cartan subalgebra are independent.
From eqs.~(\ref{rec1})--(\ref{rec4}), \eqref{k3},
\eqref{Jtres},
\eqref{accionRmenos}, \eqref{accionRmas},
\eqref{accionSmenos} and \eqref{accionSmas}
the crossed commutators between all these generators are easily computed:
\be\begin{array}{llll}\label{so32comm}
 & [J_\pm,K_\pm]=\pm R_\pm,\quad & [J_\mp,K_\pm]=\pm S_\pm,\quad & [J_3,K_\pm]=0, \qquad [J_3,K_3]=0,
\\[0.3cm]
& [J_\pm,R_\pm]=0,\quad &  [J_\mp,R_\pm]=\pm 2 K_\pm,\quad &  [J_3,R_\pm]=\pm R_\pm, 
\\[0.3cm]
& [J_\pm,S_\pm]=\pm 2 K_\pm,\quad & [J_\mp,S_\pm]=0,\quad &  [J_3,S_\pm]=\mp S_\pm ,\\[0.3cm]
&  [K_\pm,R_\pm]=0,\quad & [K_\mp,R_\pm]=\pm 2 J_\pm ,\quad & [K_3,R_\pm]=\pm R_\pm,   
\\[0.3cm]
&  [K_\pm,S_\pm]=0,\quad & [K_\mp,S_\pm]=\pm 2 J_\mp,\quad & [K_3,S_\pm]=\pm S_\pm,  
\\[0.3cm]
&  [R_\pm,S_\pm]=0,\quad &  [R_\mp,S_\pm]=0 .  & 
\end{array}\ee
From all the commutators \eqref{Kso21comm}, \eqref{so3comm},   \eqref{Rso21comm},  \eqref{Sso21comm} and 
\eqref{so32comm}   we are dealing with the noncompact real form $so(3,2)$ of $B_2$ which root diagram is
displayed in Fig.~\ref{fig_5}. 

The quadratic Casimir operator ${\cal C}^{so(3,2)}_2$ can now be evaluated to be $-5/4$
and by means of the differential form of the generators  the  equation \eqref{legendreeq}  is again obtained
\[
{\cal C}^{so(3,2)}_2 + 5/4  \equiv\,  
 X^2\left( (1-X^2) D_x^2  -2 X D_x +L(L+1)-\frac{1}{1-X^2}\;M^2\right) = 0 .
\]
The general Legendre equation is thus strictly related to   $so(3,2)$ since it  can be obtained from the quadratic invariant of $so(3,2)$ or, alternatively, from the  Casimir  of any of its  three-dimensional subalgebras.

\sect{$so(3,2)$ and  spherical harmonics}
\label{sphericalharmonicssection}

We extend now  the discussion to spherical harmonics
$Y_l^m$ that are well known to be related, for $l$ fixed, with a representation of $so(3)$. In this paper we show that the 
 SH  $\{Y_l^m\}$, for all values of $l$ and $m$, support, as the set  $\{T_l^m\}$,  the UIR with ${\cal C}_2=-5/4$ of $so(3,2)$.

Remember that spherical harmonics ${Y_l^m(\theta,\phi)}$ are 
\[
Y_l^m(\theta, \phi)\; =\; N e^{{\bf i} m \phi} \; P_l^m(x) ,
\]
where different normalizations $N$ are considered in different fields of research.
To save the correspondence with the ALP (and, in this 
way, the customary Lie algebra matrix
elements)  we fix
\be\label{norm} 
N = \frac{1}{\sqrt{2 \pi}} \sqrt{\frac{(l-m)!}{l+m)!}}.  
\ee
Hence,  the $Y_l^m(\theta,\phi)$ are:
\[
Y_l^m(\theta,\phi)\;  =\; \frac{1}{\sqrt{2 \pi}} \sqrt{\frac{(l-m)!}{(l+m)!}}\;  e^{{\bf i} m \phi} \; P_l^m(x) =
\frac{1}{\sqrt{2 \pi}}\;  e^{{\bf i} m \phi} \; T_l^m(x) ,
\]
where $x= \cos\theta$ and $\phi$ is the conjugated variable of $m$.   
Taking into account the normalization constant  (\ref{norm}),  orthogonality and completeness
of the $\{Y_l^m\}$ are \cite{WU}
\[
\int d\Omega\;\, Y_l^{m\; *}\;(\theta,\phi)\;\left(l+1/2\right)\; Y_{l'}^{m'}(\theta,\phi)\;\; 
=\;\; \delta_{l\,l'}\;  \delta_{m\,m'}\;,
\]
\[
\sum_{l\,m}\; Y_m^{m\; *}\;(\theta,\phi)\; \left(l+1/2\right)\; Y_{l}^{m}(\theta',\phi')\; =\; 
\delta\left(\cos\theta-\cos\theta'\right)\;  \delta(\phi-\phi') .
\]
With the substitutions $J_\pm \to {J_\pm}' = e^{\pm {\bf i} \phi} J_\pm$, \;   
$R_\pm \to {R_\pm}' = e^{\pm {\bf i} \phi}\, R_\pm$,\; $S_\pm \to {S_\pm}' = e^{\mp {\bf i} \phi}\, S_\pm$ and with 
 $K_\pm$ remaining invariant all the preceding results obtained for the 
$\{ T_l^m\}$ can now be easily translated to the $\{Y_l^m\}$.


\sect{Operators on $L^2$ spaces and UEA}
\label{scalprod}

We have shown in the preceding sections that both the ALP and SH are a basis of a representation of $so(3,2)$.
In this section we consider as both, the ALP and the the SH, are a basis of the $L^2$ functions defined on $(-1,1)\times \Z$ and $S^2$, respectively. Thus the vector
  space of the linear operators
acting on these $L^2$ functions  is isomorphic to the UEA of 
$so(3,2)$.  In this way the results presented in \cite{celeghini2012} 
for the one-variable square-integrable functions are extended  to these two-variable cases.

Again the discussion will be made for the  ALP since the extension to the SH is trivial.
Let us start from the separable Hilbert space   of the square-integrable functions defined
on both $E:=(-1,1) \subset {\R}$ and  $\Z$, i.e. 
${L}^2(E,\Z)$, direct sum of 
the Hilbert spaces with $m$ fixed:
${L}^2(E,\Z) =
 \bigcup_{m=-\infty}^{\infty} \;{L}^2(E,m)$.

A basis for ${L}^2(E,\Z)$ is $\{|x, m\rangle\}$ ($-1 < x <1, m \in \Z$). Orthonormality and 
completeness relations are
\be\label{xm}
\langle x,m | x', m' \rangle\; =\; \delta(x-x')\, \delta_{m\, m'} ,\qquad
\sum_{m }\, \int_{-1}^{+1} dx\, |x,m\rangle\,  \langle x,m|\; =\; {\cal I}.
\ee
As the $\{T_l^m(x)\}$ satisfy eqs.~(\ref{orthtm}) and (\ref{comptm}) 
we can now define inside the Hilbert space a new basis  $\{|l,m\rangle\}$ with $l \geq |m|$
\be\label{xxm}
|l,m\rangle := \int_{-1}^{+1} |x,m\rangle \,\sqrt{l+1/2}\; \,T_l^m(x) \,dx\;.
\ee
such that
\[
\langle l,m | l', m' \rangle\; =\; \delta_{l\,l'}\, \delta_{m\, m'}\, ,
\qquad
\sum_{ l, m}  |l,m\rangle\;   \langle l,m|\; =\; {\cal I} .
\]
Thus  the $T_l^m(x)$ play  the role of transition matrices and, as the space is real, can be written as
\be\label{xmm}
T_l^m(x) =\frac{1}{\sqrt{l+1/2}}\;  \langle x,m|l,m\rangle = 
\frac{1}{\sqrt{l+1/2}}\;\langle l,m|x,m\rangle \;. 
\ee 
In this way, in  analogy with \cite{celeghini2012}, an arbitrary vector 
$|f\rangle \, \in L^2(E,\Z)$
can be expressed as
\[
|f\rangle \;=\; \sum_{m=-\infty}^{\infty} \int_{-1}^{+1} dx\; |x,m\rangle \, \langle x,m|f\rangle \;=
\sum_{m=-\infty}^{+\infty} \;\sum_{l=|m|}^\infty |l,m\rangle \;\langle l,m|f\rangle 
\]
and we can describe $|f\rangle$  in  two alternative bases by means of the 
functions $f^m(x)$ or the succession $f_l^m$ :
\[
f^m(x):= \langle x,m|f\rangle = \sum_{l=|m|}^\infty T_l^m(x)\, f_l^m ,\qquad
f_l^m := \langle l,m|f\rangle = \int_{-1}^{+1}dx\; T_l^m(x)\, f^m(x)\; . 
\]
In particular, the completeness of the two bases determines the inner product 
\[
\langle f|g \rangle \;=   \sum_{m=-\infty}^{\infty}\;\sum_{l=|m|}^\infty  \;   g_l^m f_l^m =
\sum_{m=-\infty}^\infty \;\int_{-1}^{1}dx\, g^m(x)\, f^m(x), 
\]
as well as the Parseval identity 
\[
\sum_{m=-\infty}^{\infty}\; \sum_{l=|m|}^\infty \; [f_l^{m}]^2  = 
\sum_{m=-\infty}^{\infty}\;\int_{-1}^{1}  dx\;  [f^{m}(x)]^2 \; . 
\]

All the $L^2$ functions defined on $(E,\Z)$ can be written as
\[   
\sum_{m=-\infty}^{\infty}\;\sum_{l=|m|}^\infty  \;   T_l^m(x)\; f_l^m , 
\]
hence they
belong to the described UIR of $so(3,2)$.
The space of all linear operators that act on $L^2(E,\Z)$ is thus isomorphic the UEA of $so(3,2)$.

The SH case is similar: it can now be discussed substituting  \eqref{xm}  by the expression
\[
\langle \theta,\phi | \theta',\phi '\rangle\; =\; \delta(\cos \theta -\cos \theta')\, \delta{(\phi-\phi')} ,\qquad
 \int  d\Omega \; |\theta,\phi\rangle\,  \langle \theta,\phi|\; =\; {\cal I}.
\]
and   \eqref{xmm} by (see Ref.~\cite{WU}) 
\[
Y_l^m(\theta,\phi) = \frac{1}{\sqrt{l+1/2}}\;   \langle \theta,\phi | l m\rangle  = \frac{1}{\sqrt{l+1/2}}\;  {\langle l m | \theta, \phi \rangle}^* \,.
\]
The SH  $\{Y_l^m(\theta,\phi)\}$ as well as all $L^2$ functions defined on the sphere $S^2$ also support   the representation  with ${\cal C}=-5/4$ of $so(3,2)$ and the space of linear operators acting on the $L^2$ functions defined on the sphere is thus  isomorphic to the UEA of $so(3,2)$.
\begin{figure}
\centerline{\psfig{figure=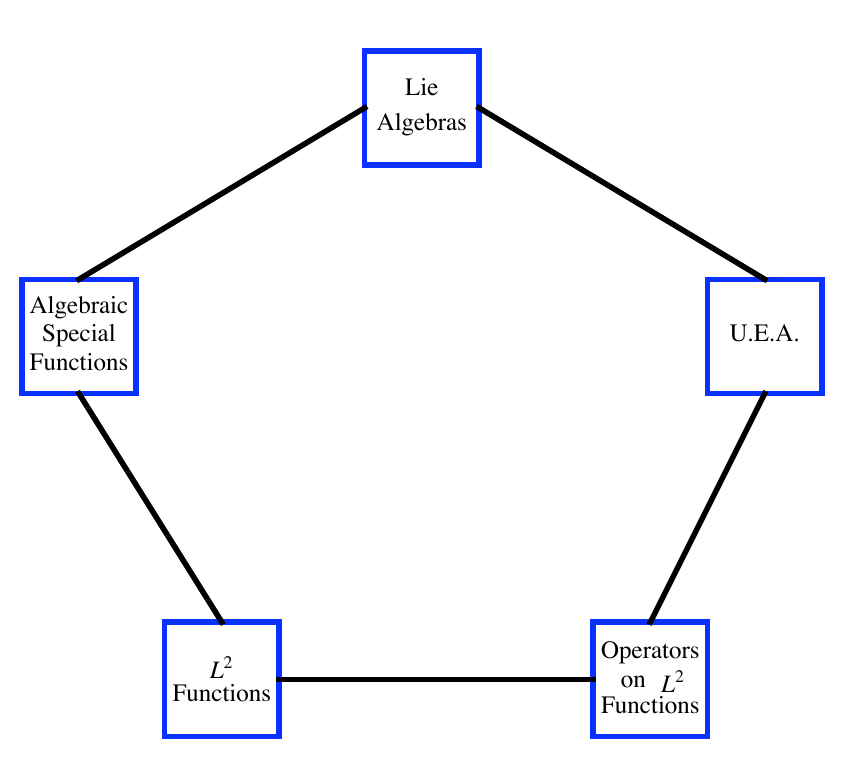,height=8.5cm}}
\caption{\small Diagrammatic resume of the philosophy of the paper.} \label{fig_6}
\end{figure}


\sect{Conclusions}\label{conclusionssection}

Following  the ideas of Talman and Truesdell we have introduced a subclass of special functions,  we call  algebraic special functions,  that support a ladder structure.  
 This excludes   elementary functions but it  includes many of  the other special functions which are familiar in physical applications. 
Their fundamental role seems to be the connection between   differential equations,  Lie algebras and  spaces of $L^2$ functions. 

Indeed they support a UIR of a  Lie algebra and, at the same time, they
are a basis in  a $L^2$ functions space. This  shows that the space of the linear operators 
acting on this $L^2$ space is  isomorphic to the UEA of the algebra.   

In particular, in this paper we discuss ALP and SH. Both are bases of the $L^2$ functions defined in the first case in the set $(-1,1)\times \Z$ and in the second one in the sphere $S^2$. They support a particular representation of $so(3,2)$ hence the operators acting on the $L^2$ functions defined in $(-1,1)\times \Z$ and in $S^2$ belong to the UEA of the $so(3,2)$.

An interesting point is that, in all the cases we have taken into account the commutators (of the algebra) and product (i.e. the factorization method) carry to the same result. It seems that it is necessary to fix both, the Lie algebra and the representation (i.e. the product) to determine the properties of differential equations we started from. 
This should imply that for algebraic special functions Casimirs of order higher of two are irrelevant in the sense that they only allow to obtain again the basic differential equation. 

 The ladder approach is suitable also for 
no simple Lie algebras. Consider, for instance, Bessel functions.
The limit  of the  ALP  to the Bessel functions \cite{bateman-project} is associated to the following multiple contractions:  1) contraction of the algebraic ladder structure, i.e., $so(3)$ to the Euclidean algebra of the plane \cite{WU}, 2)  the contraction-limit of the $L^2$ functions  from the sphere $S^2$  to the cylinder, and 3) the corresponding contraction between the operators acting on them. 

To test the scheme in a more general case we are   working now to connect Jacobi polynomials with a Lie algebra of rank three.

A pictorial description of the  ideas  behind the approach is displayed in Fig.~\ref{fig_6}.

\section*{Acknowledgments}

This work was partially supported  by the Ministerio de
Educaci\'on y Ciencia  of Spain (Projects FIS2009-09002 with EU-FEDER support),  by the
Junta de Castilla y Le\'on and by
INFN-MICINN (Italy-Spain).



\end{document}